\documentclass[12pt]{article}

\usepackage{amsfonts, amssymb, amsthm}
\usepackage{graphicx}

 \newtheorem{tw}{Theorem}
\newtheorem{de}{Definition}

\renewcommand{\theequation} {\arabic{section}.\arabic{equation}}

\def\be{\begin{equation}}
\def\ee{\end{equation}}
\def\bea{\begin{eqnarray}}
\def\eea{\end{eqnarray}}

\begin{document}
\begin{center}{ \Huge \bf  The Weyl bundle as a differentiable manifold}

\vskip0.25cm

{\bf Jaromir  Tosiek}  \\

\vskip0.25cm
{\em  Institute of Physics}\\ {\em Technical University of Lodz}\\ {\em ul. Wolczanska 219, 93-005 Lodz}\\
{\em Poland}

\vskip0.15cm

e-mail:  tosiek@p.lodz.pl

\vskip0.25cm\centerline{\today}
\begin{abstract}
Construction of an infinite dimensional differentiable manifold ${\mathbb R}^{\infty}$ not modelled on any Banach space is proposed. Definition, metric and differential structures of a Weyl algebra $(P^*_{\tt p}M[[\hbar]], \circ)$ and a Weyl algebra bundle $({\cal P^* M}[[\hbar]],\circ)$ are presented. Continuity of the $\circ$-product in the Tichonov topology is proved. Construction of the $*$-product of the Fedosov type in terms of theory of connection in a fibre bundle is explained.
\end{abstract}

\end{center}


\section{Introduction}
\setcounter{de}{0}
\setcounter{tw}{0}
Deformation quantization was born twice. First complete version of quantum mechanics in the language of classical physics appeared in the middle of the previous century, when Moyal~\cite{MO49} using  works by Weyl~\cite{WY31}, Wigner~\cite{WI32} and Groenewold~\cite{GW46} presented quantum mechanics as a statistical theory. His results were only valid for the case of $R^{2n}.$ 

For the second time  deformation quantization appeared 30 years later. Since two papers by Bayen {\it et al} \cite{bay} were published in 1978,  great interest in that version of quantum mechanics   has been observed. 

One of possible realisations of the deformation quantization programme is so called Fedosov formalism \cite{6}, \cite{7}. 
In its original version  Fedosov approach to quantum mechanics, although   based on theory of connection in a bundle, is a purely algebraic construction. A $1$-form of connection or  a $2$-form of curvature appear as objects belonging to some algebra bundle and acting on elements from that bundle via commutators. 

Presence of connection and curvature in the Fedosov's formalism framework   suggests that this topic may be treated in terms of differential geometry. As Fedosov machinery is a part of quantum theory,  its geometrization is a geometrization of deformation quantization. Profits from such treatment of a physical theory are obvious: one obtains clear definition of continuity, its is easy to represent derivation in a form covariant under some transformations etc. etc. 
Since that we find worth  reformulating the Fedosov formalism in more geometrical language.
The current paper is of our two (see \cite{my2}) devoted to this problem.

 Main role in Fedosov version of deformation quantization  is played by a Weyl bundle which is an infinite dimensional differentiable manifold. Infinite dimensional differentiable  manifolds have appeared in development of physics for several times. An example of such manifold is the Hilbert space of a quantum harmonic oscillator. But all of  known infinite dimensional differentiable manifolds {\bf are modelled on some Banach spaces} (\cite{lange}). And the Weyl bundle is not normalisable. To avoid this fundamental obstacle we propose a new look at an infinite dimensional manifold and  explain its deep physical origin.     Our considerations are  related not only with deformation quantization. They can be also useful in theory of self-dual Yang-Mills (SDYM) equations for the $*$-bracket Lie algebra  (see \cite{seba0}, \cite{seba1}).

Since that in the second section we analyse a space of infinite real series ${\mathbb R}^{\infty}$. We equip it in a metric and topological structure showing that it is a Hausdorff space and also a Fr\'{e}chet space. After that we define an atlas on that space of infinite real series and explain how to extend it to a complete atlas so we prove that ${\mathbb R}^{\infty}$ is a differentiable manifold. The space ${\mathbb R}^{\infty}$ appears as a dual space to the vector space of polynomials with real coefficients. 
 
The next part of our contribution is devoted to a Weyl algebra $(P^*_{\tt p}M[[\hbar]], \circ).$ We show that this space is a metrizable complete space modelled on the differentiable manifold ${\mathbb R}^{\infty}$ introduced in the second section. At the end of this part we prove that the $\circ$-product is continuous in the Tichonov topology.

The fourth section deals with the construction of a Weyl algebra bundle ${\cal P^* M}[[\hbar]].$ We show that the collection $\bigcup_{{\tt p} \in M}   P^*M_{\tt p}[[\hbar]]$ of the Weyl algebras is really a differentiable manifold and, moreover, is a vector bundle. 

The fifth part of our contribution is devoted to the construction of connection in the Weyl bundle. We start from a brief review of theory of connection in vector bundles and after that we introduce  symplectic connection in ${\cal P^* M}[[\hbar]].  $ In the last part of this paper, using that symplectic connection, we propose 
 abelian connection in  $P^*_{\tt p}M[[\hbar]]$ and explain  its role in  deformation quantization. In contrary to Fedosov we introduce both connections in terms of differential geometry and show that the algebraic method proposed by Fedosov is a special case (working only in Darboux atlases) of our more general treatment. Moreover, we prove that the symplectic connection is the only one  induced connection on the Weyl bundle ${\cal P^* M}[[\hbar]]  $ which can be expressed by the $\circ$-product.

In Appendix the proof of the relation between position   of an element of the Weyl algebra and its indices is presented.

As this text is concentrated  on the differential aspect of the Weyl bundle  
we do not consider its algebraic properties. Formal construction of the Weyl bundle as a bundle of elementary $C^*$ algebras is contained in \cite{ply}.
\vspace{0.5cm} 

Our paper was written by a physicist for physicists. This is the reason why we decided to  quote a lot of definitions and theorems which are well known to mathematicians. Another reason is that even such fundamental ideas like a differentiable    manifold or curvature are  defined in slightly different ways by different authors.

Bibiligraphy of the Fedosov formalism and its applications is rather wide. Since that we mention  only works the \cite{borm}--\cite{12} which represent the geometrical trend in this subject.  

Reader who finds the compendium of topology or differential geometry presented in our article unsatisfactory is pleased to look into \cite{maurin}--\cite{koba}.
Parts devoted to theory of fibre bundles are based on \cite{chern} and \cite{ward}.

Finally a short comment about notation is needed. We use the Einstein summation convention but in all formulas in which we find it necessary we put also the symbol $\sum.$ For example in the fifth section when in the same expression we have two or more sums in different intervals we use symbols of summation.   

\section{ ${\mathbb R}^{\infty}$ as a differentiable manifold}
\setcounter{de}{0}
\setcounter{tw}{0}

In this section we introduce an infinite dimensional differentiable manifold which is the natural generalisation of a space ${\mathbb R}^n, \; n \in {\cal N}.$  Such an object will be required to analyse differential properties of the Weyl bundle.

Let us  start from the $1$-dimensional ($1$--D) case.  
 A pair $({\mathbb R},\varrho )$ is a metric space with the distance defined as 
\be
\label{1t}
\varrho(x,y) \stackrel{\rm def.}{=}|x-y| \;\;\;\;\;{\rm for \;\;\; each} \;\;\; x,y \in {\mathbb R}. 
\ee
A set of balls
\[
K(c,r)=\{x \in {\mathbb R},\: |x-c|<r\}, \;\; c \in {\mathbb R}, \; r > 0, 
\] 
determines the topology ${\cal T}$ on ${\mathbb R}$ so  $({\mathbb R},{\cal T})$ is a topological space.

Let us construct a space ${\mathbb R}^{\infty}$ as the infinite Cartesian product $\Pi_{i=1}^{\infty}{\mathbb R}_i,$
where \\ $\forall_i {\mathbb R}_i= {\mathbb R}.$
In the space ${\mathbb R}^{\infty}$ the topology $\Pi_{i=1}^{\infty}{\cal T}_i$ is defined as follows. 
\begin{de} \cite{maurin}
\label{to21}
For each $x=(x^1,x^2, \ldots) \in {\mathbb R}^{\infty} $ the topological basis of  neighbourhoods of the point $x$ are all of sets ${\cal U} =\Pi_{i=1}^{\infty}{\cal W}_i, $ where 
\[
{\cal W}_i= \left\{ \begin{array}{l}
{\mathbb R} \;\;\;{\rm for \;each \;}i \; {\rm  apart \;from \;the \;finite \;number \;of \;indices,\;} \\
  \\
{\cal U}_j(x^j)\;\;\; {\rm for \; the\; rest \;of \;indices.}
\end{array}
\right.
\]
 By ${\cal U}_j(x^j)$ we mean an arbitrary $1$--D neighbourhood of $x^j \in {\mathbb R}.$ The mapping 
\[
P_i:{\mathbb R}^{\infty} \rightarrow  {\mathbb R}_i
\]
such that $P_i(x)=x^i$ is called a {\bf projection} of $\:{\mathbb R}^{\infty}$ on $  {\mathbb R}_i.$ The topology $\Pi_{i=1}^{\infty}{\cal T}_i$ is known as the {\bf Tichonov topology}. 
\end{de}
The Tichonov topology is the coarsest topology in which the projections $P_i$ are continuous.
\label{strona} 

The Tichonov topology can be also introduced by a set of seminorms. The advantage of this method is that  ${\mathbb R}^{\infty}$ becomes in a natural way a Fr\'{e}chet space so to investigate its properties we are able to use the powerful machinery of theory of Fr\'{e}chet spaces.
\begin{de} \cite{janich}
Let $V$ be a vector space over ${\mathbb R}.$ A mapping $[|\cdot|]: V \rightarrow {\mathbb R} $ is called a {\bf seminorm} if:
\begin{enumerate}
\item $\forall_{x \in V } \; [|x|] \geq 0; $
\item
$
\forall_{x \in V } \; \forall_{t \in {\mathbb R} } \; [|t \cdot x|]= |t| \cdot [|x|]$;
\item
$ \forall_{x,y \in V } \; [|x+y|] \leq [|x|] +[|y|].$ 
\end{enumerate}
An open ball in a seminorm  $[|\cdot|]$ is a set of points such that
\be
\label{ball1}
K(c,r)=\{x \in V,\: [|x-c|]<r\}, \;\; c \in V, \; \; r >0.
\ee
\end{de}

Let $[|\cdot|]_i$ be a seminorm in a vector space ${\mathbb R}^{\infty}$ defined by
\be
\label{to13}
[|\cdot|]_i: {\mathbb R}^{\infty} \rightarrow {\mathbb R}_i, \;\;\; [|x|]_i \stackrel{\rm def.}{=} |x^i|.
\ee
A set of seminorms $\{[|\cdot|]_i\}_{i \in {\cal N}}$ in ${\mathbb R}^{\infty}$ determines a topology on ${\mathbb R}^{\infty}$ in the following way. 

A set ${\cal U} \in {\mathbb R}^{\infty}$ is open in the topology compatible with seminorms $\{[|\cdot|]_i\}_{i \in {\cal N}}$ if for every $x \in {\cal U}$ there exist $i_1, \ldots, i_r \in {\cal N}$ and $\epsilon >0$ such that 
\[
K_{i_1}(x,\epsilon) \cap K_{i_2}(x,\epsilon) \cap \ldots \cap K_{i_r }(x,\epsilon) \subset {\cal U}
\]
i.e.
each point of $U$ belongs to an intersection of finite number of balls contained in $U.$ By $K_{i_l}(x,\epsilon)$ we denote an open ball (\ref{ball1}) in the seminorm $[|\cdot |]_{i_l}.$ 

Topology determined by the set of seminorms $\{[|\cdot|]_i\}_{i \in {\cal N}}$ is the same as the Tichonov topology in ${\mathbb R}^{\infty}$ introduced before ( see \cite{maurin}).
\begin{tw}\cite{janich}
A  vector space $V$ with topology ${\cal T}$ defined by the set of seminorms  \\ $\{[|\cdot|]_{z}\}_{z \in  J}$ ($J$ is countable or not)
is topological vector space $(V, {\cal T})$. It is also a Hausdorff space iff the neutral element $\Theta \in V$ is the only one vector 
such that $ \forall_{z \in J} \; [|\Theta|]_{z}=0.$
\end{tw}
Using the above theorem we conclude that
the pair $({\mathbb R}^{\infty}, \Pi_{i=1}^{\infty}{\cal T}_i)$ is a Hausdorff space.
\begin{de}\cite{janich}
A Hausdorff topological vector space $(V, {\cal T})$ is called {\bf pre-Fr\'{e}chet} if its topology is  given by a countable set of seminorms.
\end{de}
We see that the space $({\mathbb R}^{\infty}, \Pi_{i=1}^{\infty}{\cal T}_i)$ is a pre-Fr\'{e}chet vector space. All of pre-Fr\'{e}chet spaces are metrizable. The distance in 
$({\mathbb R}^{\infty}, \Pi_{i=1}^{\infty}{\cal T}_i)$ is  defined as 
\be
\label{to9}
\varrho(x,y) \stackrel{\rm def.}{=}
\sum_{i=1}^{\infty}\frac{1}{2^i}\frac{[|x-y|]_i}{1+[|x-y|]_i}=
\sum_{i=1}^{\infty}\frac{1}{2^i}\frac{|x^i-y^i|}{1+|x^i-y^i|}.
\ee
The metric (\ref{to9}) constitutes the same topology as the set of seminorms (\ref{to13}). 

There is no norm establishing the metric (\ref{to9}). This conclusion is the straightforward consequence of the fact that the maximal distance $\varrho(x,y)$ in the metric (\ref{to9}) equals $1.$ 
\begin{de}\cite{janich}
A complete pre-Fr\'{e}chet space is called a {\bf Fr\'{e}chet space }. 
\end{de}

A theorem holds
\begin{tw} \cite{maurin}
\label{maurin1}
Let $(V_i, \varrho_i), \; i=1,2, \ldots $ are metric spaces and let $v=(v_1, v_2, \ldots),$ \\ $ u=(u_1, u_2, \ldots) \in \Pi_{i=1}^{\infty}V_i $ and
\be
\varrho(v,u) \stackrel{\rm def.}{=} \sum_{i=1}^{\infty}\frac{1}{2^i}\frac{\varrho_i(v_i,u_i)}{1+\varrho_i(v_i,u_i)}.
\ee
The metric space $(\Pi_{i=1}^{\infty}V_i, \varrho )$ is complete iff all of spaces $(V_i, \varrho_i), \; i=1,2, \ldots$ are complete.
\end{tw}
The straightforward consequence of that theorem is the following one.
\begin{tw}
The pre-Fr\'{e}chet space  $({\mathbb R}^{\infty}, \Pi_{i=1}^{\infty}{\cal T}_i)$ is a Fr\'{e}chet space.
\end{tw}

\vspace{0.5cm}
Let us recall the definition of a differentiable manifold \cite{maurin}.
\begin{de}
\label{tos2}
A differentiable $n$- dimensional manifold ${\cal M}$ of class $C^r$ is a pair $({\cal M},{\cal A}),$ where ${\cal M}$ is a Hausdorff space and ${\cal A}= \{({\cal U}_z,\phi_z)\}_{z \in J}$ is a set of charts $({\cal U}_z,\phi_z)$ and 
\begin{enumerate}
\item
$ ({\cal U}_z,\phi_z)_{z \in J}$ is 
an open covering of ${\cal M}$ and $\phi_z:{\cal U}_z \rightarrow {\cal O}_z \subset {\mathbb R}^n $ are homeomorphisms  
\item
mappings
\be
\phi_{zv} \stackrel{\rm def.}{=}\phi_z \circ \phi_v^{-1}: \phi_v({\cal U}_z \cap{\cal U }_v) \rightarrow
\phi_z({\cal U}_z \cap{\cal U }_v)
\ee
are $r$-times continuously differentiable and they are called {\bf transition} functions.
\end{enumerate}
\end{de}
By ${\cal O}_z$ we denote open subsets of ${\mathbb R}^n.$
\vspace{0.5cm}

Henceforth to shorten notation instead of writing $({\mathbb R}^{\infty},\Pi_{i=1}^{\infty}{\cal T}_i)$ we will put ${\mathbb R}^{\infty}.$

The main problem in a proof that the space ${\mathbb R}^{\infty}$ is some differentiable manifold is the fact that  ${\mathbb R}^{\infty}$ is not a Banach space. To show that despite this obstacle 
 ${\mathbb R}^{\infty}$ may be treated as a differentiable manifold we propose the following consideration.
\begin{enumerate}
\item
Each element of the vector space ${\mathbb R}^{\infty}$ can be represented uniquely as an infinite series of real numbers $(x^1,x^2, \ldots).$ This suggests that there exist  atlases  on ${\mathbb R}^{\infty}$ containing only one chart $({\mathbb R}^{\infty},\phi )$ in which numbers $x^i, \; i=1,2, \ldots$ are coordinates of an arbitrary fixed point on ${\mathbb R}^{\infty}.$
\item
From physical reasons which will be explained in the next section, it is sufficient to restrict our considerations to the atlas ${\cal A}=\{({\mathbb R}^{\infty},\phi_z)\}_{z \in {J}}$  on ${\mathbb R}^{\infty}$ such that each chart $({\mathbb R}^{\infty},\phi_z) \in {\cal A}$ covers the whole space ${\mathbb R}^{\infty}$ and moreover all  mappings 
\[
\phi_{zv}=\phi_z \circ \phi_v^{-1}: {\mathbb R}^{\infty} \rightarrow {\mathbb R}^{\infty}
\]
are linear bijections. The set  of indices $J$ can be finite, countable or uncountable. Each of  bijections
 $\phi_{zv}(x^1, \ldots,x^i, \ldots)=(y^1, \ldots ,y^i, \ldots)$ 
may be written in a following form:
\be
\label{to0}
\begin{array}{ccc}
y^1& = &a_{11}x^1+a_{12}x^2+ \ldots \nonumber \\
\vdots & & \ddots  \nonumber \\
y^i& =&a_{i1}x^1+a_{i2}x^2+ \ldots  \nonumber \\
\vdots & &  \ddots 
\end{array}
\ee
where $\forall_{i,j \in {\cal N}} \;\; a_{ij} \in  {\mathbb R}.$ To ensure convergence of sums standing at the right side of the infinite system of equations (\ref{to0}) we require that for each index `$i$' only finite number of coefficients $a_{ij}$ is different from $0.$ This condition holds for every mapping $\phi_{zv}$ so especially it is true also for the transformation \\ $(y^1, \ldots,y^i, \ldots) \longrightarrow (x^1, \ldots ,x^i, \ldots)$ inverse to (\ref{to0}).
\item
Although the general definition of the derivative of the mapping $\phi_{zv}$ does not exist because of lack of a norm in ${\mathbb R}^{\infty}$  we can precisely define  partial  derivatives 
\be
\frac{\partial y^i}{\partial x^j} \stackrel{\rm def.}{=}\lim_{d \rightarrow 0}\frac{y^i(x^1, \ldots, x^j+d, \ldots)- y^i(x^1, \ldots, x^j, \ldots)}{d} \stackrel{\rm (\ref{to0}) }{=} a_{ij}.
\ee
In this infinite dimensional case we assume that the existence of all partial derivatives $\frac{\partial y^i}{\partial x^j} $ and $\frac{\partial x^j}{\partial y^i} $ (the proof for the inverse mapping is analogous) of an arbitrary range  is sufficient to treat the mapping $\phi_{zv}$ to be $C^{\infty}$- differentiable.
\item
A set of linear finite transformations of the form (\ref{to0}) constitutes a pseudogroup of transformations $\Gamma$ (see \cite{koba}). The atlas ${\cal A}=\{({\mathbb R}^{\infty}, \phi_z)\}_{z \in J}$ is compatible with the pseudogroup $\Gamma.$ Since each atlas compatible with some subgroup is contained in a {\bf unique} complete atlas of a manifold, starting from   the atlas $\{({\mathbb R}^{\infty}, \phi_z)\}_{z \in J}$ and the psedogroup $\Gamma$ we point out  the complete atlas on ${\mathbb R}^{\infty}.$
\end{enumerate}
From the construction presented in this section we conclude that ${\mathbb R}^{\infty}$ is really   differentiable manifold of a class $C^{\infty}$.

\section{The Weyl algebra}
\setcounter{de}{0}
\setcounter{tw}{0}
In this section we define and analyse some properties of a Weyl algebra. Method presented here is based on physical interpretation of this algebra. 
It is purely geometric on the contrary to algebraic way proposed by Fedosov \cite{6}, \cite{7}.
We construct the Weyl algebra by a symmetric tensor product of  spaces cotangent to some manifold. 
  
Let $({\cal M}, \omega)$ be a $2n-$D symplectic manifold, $T^*_{\tt p}M$ the cotangent space to ${\cal M}$ at a point ${\tt p}$ of ${\cal M}$ and ${\cal A}= \{({\cal U}_z,\phi_z)\}_{z \in J}$  an atlas on ${\cal M}.$
\begin{de}
The space $(T^*_{\tt p}M)^l, \; l \geq 1$  is a symmetrized tensor product of \\ $\underbrace{T^*_{\tt p}M \odot \ldots \odot T^*_{\tt p}M}_{\rm l - times}. $ It is spanned by
\be
\label{now01}
v_{\bf K_1} \odot \cdots \odot v_{\bf K_l} \stackrel{\rm def.}{=} \frac{1}{l!} \sum_{\rm all \; permutations} v_{\bf \sigma K_1} \otimes\cdots \otimes v_{\bf \sigma K_l},\ee
where $v_{\bf K_1}, \ldots,v_{\bf K_l} \in  T^*_{\tt p}M.$ For $l=0$ we put $(T^*_{\tt p}M)^0 \stackrel{\rm def.}{=}{\mathbb R}.$
\end{de}

Each element  $v \in (T^*_{\tt p}M)^l$ in a  chart $({\cal U}_z,\phi_z) \ni {\tt p}$
is uniquely represented by a sequence 
\be
\label{za1}
v=(v_{1 \ldots 1},\ldots,
v_{i_1 \ldots i_l},
\ldots, v_{2n \ldots 2n}
).
\ee
For indices  the relation holds $i_1 \leq i_2 \leq \ldots \leq i_{l-1} \leq i_{l}.$
The number of elements of the sequence (\ref{za1}) is equal to $\frac{(2n+l-1)!}{(2n-1)!\:l!}.$ 
This is the straightforward consequence of the fact that
\be
\label{dziwo}
\underbrace{\sum_{k_1=1}^m \: \sum_{k_2=k_1}^m \ldots \sum_{k_l=k_{l-1}}^m}_{\rm l- \; times} 1= \frac{(m+l-1)!}{l! (m-1)!}. 
\ee

Introducing the distance $\varrho_l(v, u) $ between two elements of $(T^*_{\tt p}M)^l$ as
\be
\label{to11}
\varrho_l(v, u) \stackrel{\rm def.}{=}
\left\{  \begin{array}{cc} |v- u| & \;\;{\rm for} \;\; l=0, \nonumber \vspace{3mm}\\ 
 \sum_{{\rm all} \;\; i_1 \leq i_2 \leq \ldots \leq i_{l-1} \leq i_{l}} |v_{i_1  i_2  \ldots  i_{l-1}  i_{l}}-u_{i_1  i_2  \ldots  i_{l-1}  i_{l}}| & \;\;{\rm for} \;\; l>0, 
\end{array} \right.
\ee
we define a metric structure in $(T^*_{\tt p}M)^l.$ The distance between $v$ and $u$ depends on the choice of the system of coordinates on the manifold ${\cal M}.$ 

A set of balls
\be
\label{to115}
K_l(v,r)=\{u \in (T^*_{\tt p}M)^l ,\: \varrho_l( v, u) < r \}, \;\; v \in (T^*_{\tt p}M)^l, \;\; r >0 
\ee 
introduces  a topology ${\cal T}_l$ on $(T^*_{\tt p}M)^l$ so  $((T^*_{\tt p}M)^l,{\cal T}_l)$ is a topological space.
Although the distance $\varrho_l(v, u)$ depends on the choice of a chart on the manifold ${\cal M}, $ the topology
${\cal T}_l$ is independent of it. This conclusion becomes obvious if we notice that the topology defined by open balls (\ref{to115}) is the same as topology established by cubes  
\[
|v_{1 \ldots 1} - u_{1 \ldots 1}| \times \cdots \times |v_{2n \ldots 2n} - u_{\:2n \ldots 2n}|.
\]
 
 Moreover 
$((T^*_{\tt p}M)^l,{\cal T}_l)$ is complete vector space.
 It  is also a differentiable manifold modelled on a Banach space $({\mathbb R}^{\frac{(2n+l-1)!}{(2n-1)!\:l!}}, ||\cdot||)$ with a norm 
\[
||x|| \stackrel{\rm def.}{=} \sum_{i=1}^{\frac{(2n+l-1)!}{(2n-1)!\:l!}}|x^i|.
\] 

\begin{de}A {\bf preWeyl vector space} $P^*_{\tt p}M$ at a point ${\tt p} \in {\cal M}$ is the direct sum \[P^*_{\tt p}M \stackrel{\rm def.}{=}\bigoplus_{l=0}^{\infty} \left( (T^*_{\tt p}M)^l \oplus (T^*_{\tt p}M)^l \right)
.\]
\end{de}
Beside $\bigoplus_{l=0}^{\infty}$ also another direct sum appears because components of tensors which are used  in physics are in general complex numbers.  The preWeyl vector space is a topological space with  the Tichonov topology. Construction of a metric and topology is analogous to that one presented in the previous section.
\be
\label{to12}
\varrho (v, u) \stackrel{\rm def.}{=} \sum_{l=0}^{\infty} \left( \frac{1}{2^{l+2}}\frac{\varrho_l(Re(v), Re(u)) }{1+ \varrho_l(Re(v), Re(u))}  
+ \frac{1}{2^{l+2}}\frac{\varrho_l(Im(v), Im(u)) }{1+ \varrho_l(Im(v), Im(u))} \right)
\ee
for each $v, u \in P^*_{\tt p}M. $ The distances $\varrho_l(Re(v), Re(u)) $ and $\varrho_l(Im(v), Im(u))$ are computed between real and imaginary parts of components of $v, u $ belonging to \\ $(T^*_{\tt p}M)^l \oplus (T^*_{\tt p}M)^l.$
 Again, although the metric (\ref{to12}) depends on the choice of a chart on ${\cal M},$ the Tichonov topology on $P^*_{\tt p}M$ is independent of it.     

Due to the theorem \ref{maurin1}   the preWeyl space is also  a Fr\'{e}chet space. 
\begin{de} \cite{5}
Let $\lambda$ be a fixed real number and $ V $ some vector space. A {\bf formal series } in the formal parameter $\lambda$ is  every expression of the form
\be
\label{1}
v[[\lambda]]= \sum_{{\bf K}=0}^{\infty}\lambda^{\bf K} v_{\bf K}, \;\;\; {\rm where} \;\; \forall_{\bf K}  \;\;\;  v_{\bf K} \in V. 
\ee
The set of formal series $v[[\lambda]]$ constitutes a vector space.  
\end{de}

Addition means vector summation of  elements of the same power of $ \lambda$ and multiplication by a scalar $a \in {\mathbb C}$ is multiplication of each vector standing on the right side of (\ref{1})  by $a$ i.e.
\be
u[[\lambda]] + v[[\lambda]] = \sum_{{\bf K}=0}^{\infty}\lambda^{\bf K} (u_{\bf K} + v_{\bf K})
\ee
and
\be
a \cdot v[[\lambda]]= \sum_{{\bf K}=0}^{\infty}\lambda^{\bf K} (a v_{\bf K}).
\ee

A vector space of formal series over the vector space $ V $ in the
formal parameter $\lambda$   we will denote by $V[[\lambda]].$ The space $V[[\lambda]]$ may be treated as a direct sum
\be
\label{11}
V[[\lambda]] = \bigoplus_{i=0}^{\infty} V_i,\;\;\; V_i=V\;\;\; {\rm for \;\;\; every \;\;\;i}.
\ee
We introduce the formal series over the preWeyl vector space as follows.
\begin{de}
 A {\bf Weyl vector space} $P^*_{\tt p}M[[\hbar]]$ is the vector space over the preWeyl vector space  $P^*_{\tt p}M$ in the formal parameter $\hbar.$
 \end{de}
 For  physical applications we usually  identify the parameter $\hbar$ with the Planck constant. Terms standing at the $k$-th power of $\hbar^k$ and belonging to the direct sum $(T^*_{\tt p}M)^l \oplus (T^*_{\tt p}M)^l$ we will denote by $v[k,l].$ Now
each element of $P^*_{\tt p}M[[\hbar]]$  may be written in the form
 \be
 \label{2}
 v= \sum_{k=0}^{\infty}  \sum_{l=0}^{\infty} \hbar^k v[k,l].
  \ee
 For $l=0 $ we put $\sum_{k=0}^{\infty} \hbar^k v[k,0], \; v[k,0] \in {\mathbb C}.$ Again using the Tichonov procedure we equip the Weyl space with a  topological structure. The Weyl space is a Fr\'{e}chet space (see  theorem \ref{maurin1}) with the metric
\be
\label{to121}
\varrho(u,v)\stackrel{\rm def.}{=}\sum_{k=0}^{\infty}\frac{1}{2^{k+1}}\frac{\varrho_k(u,
v)}{1+\varrho_k(u,v)} \:,
\ee
where by $\varrho_k(u,v)$ we understand the distance (\ref{to12}) computed between parts of $u$ and $v$ standing at  $\hbar^k.$

\begin{de} 
\cite{7}
\label{dedegree}
 The {\bf degree} ${\rm deg}( v[k,l])  $ of the component $ v[k,l] $of the  Weyl vector space $ P^*_{\tt p}M[[\hbar]]$  is the sum $2k+l.$
\end{de}

At the beginning of this paragraph we propose  following  convention - coordinates of a vector $v[k,l]$ we will denote by $v[k,l]_{i_1 \ldots i_l}.$ 
 To show that  $P^*_{\tt p}M[[\hbar]]$ is a differentiable manifold  first we introduce one chart $(P^*_{\tt p}M[[\hbar]],\phi)$ covering the whole space. The mapping $\phi$ is defined as follows:
\be
\label{to5}
\phi(v)=(Re(v[0,0]),Im(v[0,0]),Re(v[0,1]_1),Im(v[0,1]_1),\ldots ) 
\ee
Let $(\phi(v))_d$ denotes the $d-$th element of the sequence (\ref{to5}). 
Elements in this sequence are ordered according to the following rules:
\label{porzadek}
\begin{enumerate}
\item
${\rm deg} (\phi(v))_{d_1} > {\rm deg}(\phi(v))_{d_2} \Longrightarrow d_1 > d_2.$
\item 
For the same degree if the power of $\hbar$ in $(\phi(v))_{d_1}$ is higher than in $(\phi(v))_{d_2}$ then $d_1>d_2.$
\item
For the same degree and power of $\hbar$ we order elements like in (\ref{za1}) taking into account indices of a tensor .
\item
For two terms of the same degree, the same power of $\hbar$ and the same tensor indices 
the real part precedes the imaginary one.
\end{enumerate}
A formula connecting the position of $v[k,l]_{i_1 \ldots i_l}$ in the sequence (\ref{to5}) with indices $k,i_1 \ldots i_l$ is rather complicated. Reader may find it in the Appendix.

The chart $(P^*_{\tt p}M[[\hbar]],\phi)$  is determined by the choice of a chart on the symplectic manifold ${\cal M}$ because  numbers $(\phi(v))_d $ are components of tensors in a natural basis given by coordinates on ${\cal M}.$ Let us cover the Weyl space with an atlas ${\cal A}= \{({\cal U}_z,\phi_z)\}_{z \in J}$ consisting of all natural charts. Each of mappings $\phi_z$ satisfies the ordering rule (\ref{to5}) and covers the whole Weyl space.
\begin{enumerate}
\item 
$\forall_{z \in J} \phi_{z}$ are homeomorphisms $(P^*_{\tt p}M[[\hbar]],\phi) \rightarrow {\mathbb R}^{\infty}.$
\item 
mappings $\phi_{zv}= \phi_z \circ \phi_v^{-1}:{\mathbb R}^{\infty} \rightarrow {\mathbb R}^{\infty}$ are linear bijections of the kind (\ref{to0}). Moreover, each arbitrary fixed element  $(\phi_z(v))_i$  depends linearly only on terms $(\phi_v(v))_j$  characterized by the same power of $\hbar,$ the same tensor range and belonging to the same real or imaginary part of $v.$ Thus we conclude that all partial derivatives $\frac{\partial (\phi_z(v))_j}{\partial (\phi_v(v))_i}$ exist. Moreover, for a fixed $j$ only finite number of those partial derivatives do not vanish. 
\end{enumerate}
Taking into account facts presented above we say that the Weyl space $P^*_{\tt p}M[[\hbar]]$ is a differentiable manifold.

 For physical reasons (see \cite{7}) the Weyl space may be equipped with a structure of an algebra. 
Let $X_{\tt p} \in T_{\tt p}M$ be some fixed vector from the  space $ T_{\tt p}M$    tangent to ${\cal M}$ at a point ${\tt p}.$ Components of $X_{\tt p}$ in the natural basis $\{\frac{\partial}{\partial q^1}, \ldots,
\frac{\partial}{\partial q^{2n}} \}$  we denote by $ X_{\tt p}^i .$ 
 It is clear that for every $v[k,l] \in P^*_{\tt p}M[[\hbar]]$ the acting 
\[
v[k,l](\underbrace{X_{\tt p}, \ldots,X_{\tt p}}_{\rm l - times})= v[k,l]_{i_1 \ldots i_l}X_{\tt p}^{i_1} \cdots X_{\tt p}^{i_l} 
\]
is a complex number and we can treat $v[k,l](\underbrace{X_{\tt p}, \ldots,X_{\tt p}}_{\rm l - times})$
as a polynomial of the $l-$th degree in components of the vector $X_{\tt p}.$ 

Thus 
 elements of the Weyl space $P^*_{\tt p}M[[\hbar]]$ are mappings
\[
 v(X_{\tt p} ): {\mathbb R}^{2n} \rightarrow {\mathbb C}[[\hbar ]], 
 \]
 \be
 \label{now001}
    v(X_{\tt p} ) \stackrel{\rm def.}{=}\sum_{k=0}^{\infty} \sum_{l=0}^{\infty} \hbar^k v[k,l]_{i_1 \ldots i_l} X_{\tt p}^{i_1} \cdots X_{\tt p}^{i_l}. 
\ee
A symbol ${\mathbb C}[[\hbar ]]$ denotes a space of formal series over ${\mathbb C}$.
\vspace{0.5cm} 

Let us go to the definition of $\circ$-product in the Weyl space.
By a derivative $\frac{\partial v}{\partial X_{\tt p}^{i}} $ we understand  formal sum (ordered by powers of $\hbar^k$) of partial derivatives of polynomials $v[k,l]_{i_1 \ldots i_l} X_{\tt p}^{i_1} \cdots X_{\tt p}^{i_l}.$ Derivation $\frac{\partial }{\partial X_{\tt p}^{i}}$ does not influence on powers of $\hbar.$
The definition of derivation presented here is formal because in the space ${\mathbb C}[[\hbar ]]$ a norm is not defined. 
  
\begin{de}   \cite{7}
The product $\circ: P^*_{\tt p}M[[\hbar]] \times P^*_{\tt p}M[[\hbar]]\rightarrow P^*_{\tt p}M[[\hbar]]$ of two elements $
v, u \in P^*_{\tt p}M[[\hbar]]$ is such element  $ w \in P^*_{\tt p}M[[\hbar]]$ that for each $X_{\tt p} \in T_{\tt p}M$ the equality holds
\[
w(X_{\tt p})=  v(X_{\tt p}) \circ u(X_{\tt p})=
\]
\be
\label{6}
\stackrel{\rm def.}{=}  \sum_{t=0}^{\infty} \frac{1}{t!}\left(\frac{i\hbar}{2}\right)^t\omega^{i_1 j_1} \cdots \omega^{i_t j_t} \:\frac{\partial^t \; v(X_{\tt p}) }{\partial X^{i_1}_{\tt p}\ldots\partial X^{i_t}_{\tt p} }\:\frac{\partial^t \; u(X_{\tt p}) }{\partial X^{j_1}_{\tt p}\ldots\partial X^{j_t}_{\tt p} }.
\ee
\end{de}
The pair $(P^*_{\tt p}M[[\hbar]],\circ) $ is a noncommutative associative algebra called the {\bf Weyl algebra}. By $\omega^{ij}$ we understand components of the tensor inverse to the symplectic form in a point ${\tt p}$ i.e. the relation holds
\[
\omega^{ij}\omega_{jk}= \delta^i_k.
\]

Here we mention some properties of the $\circ$-product in $(P^*_{\tt p}M[[\hbar]],\circ). $ More information can be found in  \cite{7}, \cite{my2}, \cite{my1}.  It is worth  emphasizing that the first of presented properties becomes clear thanks to   our geometrical approach to the Weyl algebra. 
\begin{enumerate}
\item 
The $\circ$-product is independent of the chart.
\item
The $\circ$-multiplication is associative but nonabelian.
\item
\label{to20}
$\forall_{v, u \: \in \:(P^*_{\tt p}M[[\hbar]],\circ)} $ the relation holds
$
\;\;\; \deg(v \circ u)= \deg(v) + \deg(u).
$
\end{enumerate}

Let us show continuity of the $\circ$-product in the Tichonov topology in $(P^*_{\tt p}M[[\hbar]],\circ).$ Analogously to the formula (\ref{to13}) in an arbitrary fixed chart $(P^*_{\tt p}M[[\hbar]],\phi)$ we introduce  seminorms  as
\be
\label{to14}
\forall_{v \in (P^*_{\tt p}M[[\hbar]],\circ)} \; \forall_{i \in {\cal N}}\;  [|v|]_i \stackrel{\rm def.}{=} |(\phi(v))_i|.
\ee

 The metric  (\ref{to121}) is now expressed by seminorms $\{[|\cdot|]_i\}_{i \in {\cal N}}$ as (compare with  (\ref{to9}))
\be
\varrho(v,u)= \sum_{i=1}^{\infty}\frac{1}{2^i}\frac{[|v-u|]_i}{1+[|v-u|]_i}.
\ee
Let us consider two sequences  $\{v_{\bf K} \}_{{\bf K} \in {\cal N}},\{ u_{\bf J} \}_{{\bf J} \in {\cal N}}$ elements $v_{\bf K}, u_{\bf J} \in (P^*_{\tt p}M[[\hbar]],\circ)$ such that 
\be
\label{to15}
 \lim_{{\bf K} \rightarrow \infty} v_{\bf K}= \lim_{{\bf J} \rightarrow \infty} u_{\bf J} = \Theta.
\ee
This relation  implies that for every $i \in {\cal N}$ 
\be
\label{to16}
\lim_{{\bf K} \rightarrow \infty} [|v_{\bf K}|]_i= 
\lim_{{\bf J} \rightarrow \infty} [|u_{\bf J}|]_i=0.
\ee
The $\circ$-product is continuous if for every two  sequences $\{v_{\bf K}\}_{{\bf K} \in {\cal N}}, \{u_{\bf J}\}_{{\bf J} \in {\cal N}}$ fulfilling the condition (\ref{to15}) and for every two $a, b \in (P^*_{\tt p}M[[\hbar]],\circ)$  the equality holds
\be
\label{to17}
\lim_{{\bf J},{\bf K} \rightarrow \infty}\left( (a + v_{\bf K}) \circ (b + u_{\bf J}) \right) =a \circ b. 
\ee
Formula (\ref{to17}) is equivalent to the system of equations 
\bea
\label{to18}
& \lim_{{\bf K} \rightarrow \infty} v_{\bf K} \circ b =\Theta,&  \\
\label{to18a}
& \lim_{{\bf J} \rightarrow \infty} a\circ u_{\bf J} =\Theta,&   \\
\label{to18b}
& \lim_{{\bf J},{\bf K} \rightarrow \infty} v_{\bf K} \circ u_{\bf J} =\Theta.& 
\eea
Let us consider an  element  $(v_{\bf K} \circ b )[r,l]_{i_1 \ldots i_l}$. Its degree equals $2r+l.$ From  the definition of the $\circ$-product (\ref{6}) and its 3rd property (look at page \pageref{to20}) we see that $(v_{\bf K} \circ b )[r,l]_{i_1 \ldots i_l}$ is a finite linear combination of components of vectors $v_{\bf K}$ such that $\deg v_{\bf K} \leq 2r+l.$ Since that for ${\bf J} \rightarrow \infty $ the equality (\ref{to18}) holds. Using the same method we prove relations (\ref{to18a}) and (\ref{to18b}). 
\rule{2mm}{2mm}

\section{The Weyl bundle}
\setcounter{de}{0}
\setcounter{tw}{0}
Until now we have worked with the Weyl algebra 
 $ ( P^*M_{\tt p}[[\hbar]], \circ)$ at an arbitrary fixed point ${\tt p}$ belonging to the symplectic manifold ${\cal M}$. Now we are going to analyse a collection of Weyl algebras taken for all  points of ${\cal M}.$
\begin{de}
\label{tos0}
 A {\bf Weyl bundle} is a triplet $({\cal P^*M}[[\hbar]],\pi, {\cal M}),$ where 
\be
{\cal P^*M}[[\hbar]] \stackrel{\rm def.}{=} \bigcup_{{\tt p} \in {\cal M}}   P^*M_{\tt p}[[\hbar]]
\ee
is a differentiable manifold called a total space, ${\cal M}$ is a base space  and  \\ $\pi:{\cal P^*M}[[\hbar]] \rightarrow {\cal M}$  a projection.
\end{de}
Elements of the Weyl bundle ${\cal P^*M}[[\hbar]]$  can be thought of as the pairs $({\tt p}, v),$ where $v \in ( P^*M_{\tt p}[[\hbar]],\circ).$
The projection $\pi$ assigns  a point ${\tt p}$ to the pair  $({\tt p}, v). $ 
\vspace{0.5cm}

The Weyl algebra $(P^*_{\tt p}M[[\hbar]], \circ)$ considered in the previous section is related to the point ${\tt p} \in {\cal M}.$ To define a fibre in the Weyl bundle ${\cal P^*M}[[\hbar]]$ it is required to introduce the $\circ$-product in the topological vector space ${\mathbb R}^{\infty}. $ However,
the explicit form of a product $ u \circ v,$ where $u,v \in {\mathbb R}^{\infty}$ are of the form (\ref{to5}), is complicated and useless for practical purposes. But in fact  it is sufficient to be aware that the algebra $({\mathbb R}^{\infty}, \circ)$ exists.

The definition \ref{tos0} contains a statement that 
the Weyl bundle ${\cal P^*M}[[\hbar]]$ is a $C^{\infty}$ differentiable manifold. To prove this fact we define a topology on ${\cal P^*M}[[\hbar]]$  first.

Let ${\cal A}=\{({\cal U}_z,\phi_z )\}_{z \in J}$ be an atlas on the  ${\cal  M}.$ By the definition
\[
P' \:\stackrel{\rm def.}{=}\bigcup_{z \in J} \left( {\cal U}_z \times {\mathbb R}^{\infty} \times \{z\}\right) \subset {\cal M} \times {\mathbb R}^{\infty} \times J.
\]
The set $J$ is equipped with the discrete topology. The topology in  ${\cal M} \times {\mathbb R}^{\infty} \times J$ is constructed according to  the definition \ref{to21} so the Cartesian product ${\cal M} \times {\mathbb R}^{\infty} \times J$ has a Tichonov topology. Since $\bigcup_{z \in J} \left( {\cal U}_z \times  {\mathbb R}^{\infty} \times \{z\} \right)$ is a subset of 
${\cal M} \times {\mathbb R}^{\infty} \times J,$ we equip  it with the induced topology. 

Now in $P'$ we establish the equivalence relation 
\be
\label{to23}
({\tt p},v, z) \sim ({\tt p}',v', d) \;\;\; {\rm iff}
\ee
\begin{enumerate}
\item
$
{\tt p}={\tt p}', 
$
 \item
$v' $ is the image of $v$ in the mapping induced by the change of coordinates \\ $({\cal U}_z, \phi_z) \rightarrow ({\cal U}_d, \phi_d)$ on ${\cal M}.$ It is important to remember that  the tensor $\omega^{ij}$ also transforms under that change of coordinates. 
\end{enumerate}
 ${\cal P^*M}[[\hbar]] \stackrel{\rm def.}{=}P'/ \sim \:$ posesses the quotient topology so the Weyl bundle is a topological space.

 Let us consider two projections:
 \begin{enumerate}
 \item
 a canonical projection 
 $G: P' \rightarrow P'/ \sim $  defined as
 \be
 \forall_{({\tt p},v,z) \in P' } \;G(({\tt p},v,z)) = [({\tt p},v,z)].
 \ee
Since the topology on ${\cal P^*M}[[\hbar]]$ is the quotient one, the canonical projection $G$ and its inverse $G^{-1}$ are  continuous mappings \cite{kura}. A symbol $[\;\cdot \;] $ denotes the equivalence class;
\item
a projection
$L: P' \rightarrow {\cal M}$ fulfilling the equation
\be
\forall_{({\tt p},v,z) \in P' } \;L(({\tt p},v,z)) = {\tt p}.
\ee
In  the topology of a Cartesian product on  ${\cal M} \times {\mathbb R}^{\infty} \times J $  a projection \\ $Pr: {\cal M} \times {\mathbb R}^{\infty} \times J \rightarrow {\cal M} $ defined as
\[
\forall_{({\tt p},v,z) \in {\cal M} \times {\mathbb R}^{\infty} \times J } \;Pr(({\tt p},v,z)) = {\tt p}
\]
is continuous from the definition of the Tichonov topology. The projection $L$ is nothing but $Pr|_{P'}$ and the topology on $P'$ has been induced from ${\cal M} \times {\mathbb R}^{\infty} \times J$ so the mapping $L$ is continuous. 
\end{enumerate}
Projections $G$ and $L$ preserve the point ${\tt p} $
so we can draw a commutative diagram
\setlength{\unitlength}{1ex}
\begin{picture}(30,20)
\put(29,15){$P'$}
\put(32,15.5){\vector(1,0){22}}
\put(43,17){\footnotesize $ L$}
\put(55,15){${\cal M}$}
\put(32,14){\vector(1,-1){10}}
\put(38,9){\footnotesize $G$}
\put(39,1){${\cal P^*M}[[\hbar]]$}
\put(44,4){\vector(1,1){10}}
\put(47,9){\footnotesize $\pi$}
\end{picture}

The mapping $\pi \stackrel{\rm def.}{= } L \circ G^{-1} $ as a product of two continuous mappings is also continuous.
The equality holds
\be
\label{tak}
G({\cal U}_z \times  {\mathbb R}^{\infty} \times \{z\})= \pi^{-1}({\cal U}_z).
\ee
Relation (\ref{tak})
constitutes a bijection between $ {\cal U}_z \times  {\mathbb R}^{\infty}  \times \{z\}$ and $\pi^{-1}({\cal U}_z).$ Moreover $G$ and $G^{-1}$ are continuous mappings  so we conclude that
 \be
 \label{raq}
 G|_{ {\cal U}_z \times  {\mathbb R}^{\infty} \times \{z\}}: {\cal U}_z \times  {\mathbb R}^{\infty}  \times \{z\} \rightarrow \pi^{-1}({\cal U}_z)
 \ee
  is a  homeomorphism. 
 
 Next step is to prove that ${\cal P^*M}[[\hbar]]$ is a Hausdorff space. Let ${\tt q}_1, {\tt q}_2 \in {\cal P^*M}[[\hbar]]$ and $\pi({\tt q}_1) \neq \pi({\tt q}_2).$ The manifold ${\cal M}$ is a Hausdorff space so we can always choose two neighbourhoods ${\cal V}_1,{\cal V}_2  $ such that $\pi({\tt q}_1) \in {\cal V}_1,\pi({\tt q}_2) \in {\cal V}_2 $  and ${\cal V}_1 \cap {\cal V}_2 = \emptyset.$ 
 
Let us introduce  the identification 
 $Id: {\cal U}_z \times {\mathbb R}^{\infty} \times \{z\} \cong {\cal U}_z \times {\mathbb R}^{\infty}.$ Superposition of two mappings
 \be
 \label{-1}
 Id \circ (G|_{ {\cal U}_z \times  {\mathbb R}^{\infty} \times \{z\}})^{-1} : \pi^{-1}( {\cal U}_z) \rightarrow {\cal U}_z \times {\mathbb R}^{\infty} 
 \ee
 is a homeomorphism so $\pi^{-1}({\cal V}_1)$ and $\pi^{-1}({\cal V}_2)$ are neighbourhoods of  ${\tt q}_1, {\tt q}_2$ respectively and, moreover $\pi^{-1}({\cal V}_1) \cap \pi^{-1}({\cal V}_2)= \emptyset.$ 
 
 In case ${\tt q}_1,  {\tt q}_2 \in {\cal P^*M}[[\hbar]],{\tt q}_1 \neq  {\tt q}_2 $ and $\pi({\tt q}_1) = \pi({\tt q}_2)$ we have ${\tt q}_1,  {\tt q}_2 \in \pi^{-1}({\cal U}_z)$ for some $z \in J.$ But  from the fact that (\ref{-1}) is homeomorphism we conclude that  $\pi^{-1}({\cal U}_z)$ is a Hausdorff space so there exist separate neighbourhoods of ${\tt q}_1$ and $  {\tt q}_2.$
 
 So it has been proved  that the Weyl bundle is  a Hausdorff space.
 
 The Cartesian product of the open subset ${\cal U}_z $ and the fibre ${\mathbb R}^{\infty}$  is homeomorphic to ${\cal O}_z \times {\mathbb R}^{\infty},$ where ${\cal O}_z $ is an open subset of ${\mathbb R}^{2n}.$ We denote that homeomorphism by \\ $\varsigma_z:{\cal U}_z \times {\mathbb R}^{\infty} \rightarrow {\cal O}_z \times {\mathbb R}^{\infty} . $ The mapping $\varsigma_z $ is determined by a chart $({\cal U}_z, \phi_z).$ Namely
 \be
 \varsigma_z ({\tt p},v)= (\phi_z({\tt p}), T_{\phi_z}(v)).
 \ee
 The symbol $T_{\phi_z}$ denotes taking natural components of $v.$ Remember that also the tensor $\omega^{ij}$  from the definition (\ref{6}) transforms under $\varsigma_z.$ 
 The homeomorphism
 \[
 \varphi_z: \pi^{-1}({\cal U}_z) \rightarrow {\cal O}_z \times {\mathbb R}^{\infty},
 \]
\be
\varphi_z \stackrel{\rm def.}{=} \varsigma_z \circ  Id \circ G^{-1}
\ee
establishes a differential structure on the Weyl bundle ${\cal P^*M}[[\hbar]].$ Indeed 
an atlas on ${\cal P^*M}[[\hbar]]$ is  the set of charts $\{(\pi^{-1}({\cal U}_z), \varphi_z)\}_{z \in J}.$
Mappings $\varphi_z \circ \varphi_v^{-1} $ are transition functions and they are $C^{\infty}$-differentiable in a sense that all of their partial derivatives exist. 

 Thus we conclude that ${\cal P^*M}[[\hbar]] $ is a $C^{\infty}$-class infinite dimensional differentiable manifold. 
\vspace{0.5cm}

The Weyl bundle is an example of a vector bundle. Apart from elements described above the  definition of  a vector bundle (for details see \cite{10}) it  constists  of  a structure group ${\cal G}$ being a Lie group and local trivialisations. 

If be $GL(2n,{\mathbb R} )$ we denote the group of real automorphisms of the cotangent space $T^*_{\tt p}M,$ the structure group of the fibre is
\be
{\cal G} \stackrel{\rm def.}{=} \bigoplus_{z=0}^{\infty} \bigoplus_{k=0}^{[\frac{z}{2}]} \left( \underbrace{GL(2n,{\mathbb R})\otimes \ldots  \otimes GL(2n,{\mathbb R})}_{\rm (z-2k)- times} \oplus \underbrace{GL(2n,{\mathbb R})\otimes \ldots  \otimes GL(2n,{\mathbb R})}_{\rm (z-2k)- times} \right).
\ee
Moreover, when the element $v_{i_1 i_2}$ transforms under the element \\ $g \in GL(2n,{\mathbb R}) \otimes GL(2n,{\mathbb R})$ then $\omega^{ij}$ transforms under $g^{-1} \in GL(2n,{\mathbb R}) \otimes GL(2n,{\mathbb R}).$ 

Mappings $Id \circ (G|_{ {\cal U}_z \times  {\mathbb R}^{\infty} \times \{z\}})^{-1} $ (see (\ref{raq})) are local trivialisations of the Weyl bundle because they map  $ \pi^{-1}( {\cal U}_z)$ onto the direct product  $ {\cal U}_z \times {\mathbb R}^{\infty}. $
 
Since the fibre ${\mathbb R}^{\infty}$ is not only a vector space but also an algebra, the Weyl bundle is an example of the algebra bundle.

Knowing that the Weyl bundle is a differentiable manifold  we can easily define smooth sections of it or introduce a paralell transport on ${\cal P^*M}[[\hbar]]$. Physical application of those quantities will be explained in the last part of our contribution.  

%%%%%%%%%%%%%%%%%%%%%%%%%%%%%%%%%%%%%%%%%%%%%%%%%%%%%%%%%%%%%%%%%%%%%%%%%%%%%%%%%%%%%%%%%%%%%%%%%%%%%%%%%%%%%%%%%%%%%%%%%%%
%%%%%%%%%%%%%%%%%%%%%%%%%%%%%%%%%%%%%%%%%%%%%%%%%%%%%%%%%%%%%%%%%%%%%%%%%%%%%%%%%%%%%%%%%%%%%%%%%%%%%%%%%%%%%%%%%%%%%%%%%%%
%%%%%%%%%%%%%%%%%%%%%%%%%%%%%%%%%%%%%%%%%%%%%%%%%%%%%%%%%%%%%%%%%%%%%%%%%%%%%%%%%%%%%%%%%%%%%%%%%%%%%%%%%%%%%%%%%%%%%%%%%%%
\section{Connections in the Weyl bundle}
Now we are ready to present the construction of a connection in the Weyl bundle. This construction plays crucial role in physical applications of mathematics contained in our contribution. As before we begin with some general definitions and later apply them to the Weyl algebra bundle. 
\begin{de}\cite{chern}
\label{kono}
Suppose $( E, \pi,{\cal M})$ is a vector bundle over a manifold ${\cal M} $ and $C^{\infty}(E)$ is a set of smooth sections of $E$ over ${\cal M}$. An {\bf exterior covariant derivative} on the bundle $(E, \pi,{\cal M})$ is a map
\[
\tilde{\partial}:C^{\infty}(E) \longrightarrow C^{\infty}({\cal T^*M} \otimes E),
\] 
which satisfies the following conditions:
\begin{enumerate}
\item 
for any $u,v \in C^{\infty}(E)$
\be
\label{kolin}
\tilde{\partial}(u + v)= \tilde{\partial}(u) + \tilde{\partial}(v),
\ee 
\item
for any $v \in C^{\infty}(E)$ and any $f \in C^{\infty}({\cal M})$
\be
\label{kol1}
\tilde{\partial}(f v )= df \otimes v + f \cdot \tilde{\partial}v.
\ee
\end{enumerate} 
\end{de}
\underline{Remarks}
\begin{enumerate}
\item
The map $\tilde{\partial}$ is also called connection (see  \cite{6}, \cite{7}, \cite{chern}, \cite{ward}). We keep convention used in our earlier papers and by `connection' understand only a form determining exterior covariant derivative of a local frame field (see formula (\ref{k1})).
\item
By ${\cal T^*M}$ the cotangent bundle over the manifold ${\cal M}$ is denoted. 
\end{enumerate} 

Let us introduce a new symbol. $\Lambda^p({\cal M})$ is a bundle of $p$-forms over ${\cal M}.$ 
Smooth sections of the tensor product $E \otimes \Lambda^p({\cal M})$ are known as $p$-forms on ${\cal M}$ of values in $E$ or $E$-valued $p$-forms. Since $C^{\infty}(E \otimes \Lambda^p({\cal M}))$ is a module over $C^{\infty}({\cal M}),$ there is a module isomorphism (\cite{ward})
\be
\label{isom}
C^{\infty}(E) \otimes C^{\infty}(\Lambda^p({\cal M})) \longrightarrow C^{\infty}(E \otimes \Lambda^p({\cal M}))
\ee
which we shall denote by $v \otimes f \rightarrow f \cdot v$ or simply $f v.$
It means the the second condition from the definition \ref{kono} takes the form
\be
\label{k15}
\tilde{\partial}(f v )= df \cdot v + f \cdot \tilde{\partial}v.
\ee

We are ready to find more `operational' form of the map $\tilde{\partial}.$
Let a system of vectors $e_1, e_2, \ldots, e_g $ (in general an infinite one) constitute a basis of the fibre $E_{\tt p}$ at a point ${\tt p} \in {\cal M}.$ 
Suppose that 
 a matrix of base vectors is given by
\be
e= [e_1 \; e_2 \; \ldots\; e_g].
\ee
Moreover let 
\be
\label{k5}
A=
\left[
\begin{array}{cccc}
a^1_1 & a^1_2 & \ldots & a^1_g \\
a^2_1 & a^2_2 & \ldots & a^2_g \\
 \vdots & \vdots & \ddots & \vdots \\
a^g_1 & a^g_2 & \ldots & a^g_g
\end{array}
\right]
\ee
be a nonsingular real matrix. In case when $g= \infty$ it is supposed that in each row only finite number of terms is different from $0.$ The same assumption is true for the inverse matrix $A^{-1}.$ Each system of versors $e' $ such that
\be
\label{k12}
e'=e \cdot A
\ee
is also a basis of $E_{\tt p}.$
A vector $v \in E_{\tt p}$ in the basis $e$ is represented by the  $1$-column matrix
\be
\label{k3}
v=
\left[
\begin{array}{c}
v^1 \\ v^2 \\ \vdots \\v^g
\end{array}
\right].
\ee
 Under the linear map determined by the matrix $A$ the transformation rule for  $v$ is given by the formula
\be
\label{k13}
v'=A^{-1}\cdot v.
\ee
 
Having given the basis  $e$ of the fibre at the point ${\tt p}$ we need to propagate it smoothly on the whole neighbourhood $ {\tt p} \in {\cal U} \subset {\cal M}$ of the point ${\tt p}.$
To do it we  choose smooth sections ${\bf e_1}, {\bf e_2}, \ldots, {\bf e_g} $ of the bundle $E$ over ${\cal U}$ in such manner  that in each point $ {\tt q} \in {\cal U}\: \forall_{i} \:{\bf e_i}|_{\tt q} =e_i.$ That set of sections constitutes a {\bf local frame field} of $E$ on ${\cal U}.$ At every point ${\tt q} \in {\cal U}$ the system of vectors $dq^j \otimes  e_i, 1\leq j \leq 2n, 1 \leq i \leq g \;$
forms a basis of $T^*_{\tt p}M \otimes E.$
Since an exterior covariant derivative $D{\bf e_i}$ is a local section of ${\cal T^*M} \otimes E,$
we can write
\be
\label{k1}
\tilde{\partial}{\bf e_i}= \sum_{j=1}^{2n} \sum_{k=1}^{g} \Gamma^{k}_{ij}dq^j \otimes {\bf e_k}.
\ee 
Coefficients $\Gamma^{k}_{ij}$ are smooth real functions on ${\cal U}$ and they are called  {\bf connection coefficients  } on $E$ over ${\cal U}.$ It is always possible to propagate a connection on the whole bundle (see \cite{chern}).  

Introducing the  connection matrix
\be
\label{k2}
\varpi^{k}_i \stackrel{\rm def.}{= } \sum_{j=1}^{2n} \Gamma^{k}_{ij}dq^j,
\ee
\be
\label{k4}
\varpi =
\left[
\begin{array}{ccc}
\varpi^1_1 & \varpi^1_2 & \ldots  \\
\varpi^2_1 & \varpi^2_2 & \ldots  \\
 \vdots & \vdots & \ddots 
\end{array}
\right]
\ee
we see that 
\[
\tilde{\partial}{\bf e_i}=\sum_{k=1}^{g}  \varpi^{k}_i  \otimes {\bf e_k} \stackrel{\rm (\ref{isom})}{=} \sum_{k=1}^{g}  \varpi^{k}_i 
{\bf e_k}
\]
or, equivalently, in terms of matrices
\[
\tilde{\partial}{\bf e}=  \varpi \cdot {\bf e}.
\]
It means that the exterior covariant derivative of a section $v \in C^{\infty}( E)$ over ${\cal U}$ equals
\[
\tilde{\partial}v= D(\sum_{i=1}^{g} v^i {\bf e_i})\stackrel{\rm (\ref{kolin}), (\ref{kol1})}{=} \sum_{i=1}^{g} dv^i \otimes {\bf e_i} + \sum_{i,k=1}^{g} v^i \varpi_i^k
\otimes {\bf e_k} \stackrel{\rm (\ref{isom})}{=}
\]
\be
\label{k14}
\sum_{i=1}^{g} dv^i  {\bf e_i} + \sum_{i,k=1}^{g} v^i \varpi_i^k {\bf e_k}.
\ee

As it can be proved (\cite{ward}) the connection matrix $\varpi $ transforms according to the rule
\be
\label{k7}
\varpi'= A^{-1} \cdot dA + A^{-1} \cdot \varpi \cdot A.
\ee 
Till now we were able to differentiate only vectors from the bundle $E.$ But it is possible to extend the operation $\tilde{\partial}$ to $p$-forms with values in $E.$
\begin{tw}\cite{ward}
There is a unique operator 
\[
\partial: C^{\infty}(E \otimes \Lambda^p({\cal M})) \longrightarrow C^{\infty}(E \otimes \Lambda^{p+1}({\cal M}))
\]
satisfying
\begin{enumerate}
\item
${\rm for \; all}\;\;\;f \in C^{\infty}( \Lambda^{q}({\cal M})), \; v \in C^{\infty}(E \otimes \Lambda^p({\cal M}))$
\be
\partial (f \wedge v)= df \wedge v + (-1)^q f \wedge \partial v,
\ee
\item
for $v \in C^{\infty}(E)$
\be
\tilde{\partial}v =\partial v.
\ee
\end{enumerate}
\end{tw}
It is easy to check that for $v^i \in C^{\infty}(\Lambda^p({\cal M}))$ and ${\bf e_i} \in C^{\infty}(E)$
\[
\partial v=\partial(\sum_{i=1}^g v^i {\bf e_i})= \sum_{i=1}^g (dv^i + \sum_{j=1}^g \varpi^i_j \wedge v^j){\bf e_i},
\]
what is usually written as
\be
\label{k16}
\partial v=dv + \varpi \wedge v.
\ee
This defines  a sequence of mappings
\[
C^{\infty}(E) \stackrel{\partial}{\longrightarrow} C^{\infty}(E \otimes \Lambda({\cal M}))
\stackrel{\partial} {\longrightarrow} C^{\infty}(E \otimes \Lambda^2({\cal M}))
\stackrel{\partial} {\longrightarrow}\ldots 
\]
Let us consider the second exterior covariant derivative 
\[
\partial^2: C^{\infty}(E) \longrightarrow C^{\infty}(E \otimes \Lambda^2({\cal M})).
\]
After short computations we see that
\[
\partial^2 v=\sum_{i=1}^g ( \sum_{j=1}^g d\varpi^i_j \wedge v^j + \sum_{j,k=1}^g \varpi^i_k \wedge \varpi^k_j \wedge v^j)\, {\bf e_i} 
\]
or in a compact form
\be
\label{k17}
\partial^2 v=(d\varpi +\varpi \wedge \varpi) \wedge v.
\ee
\begin{de}
The {\bf curvature } of the connection $\varpi $ on the vector bundle $(E,\pi, {\cal M})$ is a $2$-form
\be
\label{k8}
{\cal R} \stackrel{\rm def.}{=}d \varpi + \varpi \wedge \varpi.
\ee
\end{de}
The transformation rule for curvature under the linear transformation (\ref{k5})  is expressed by a formula
\be
\label{k9}
{\cal R}'= A^{-1} \cdot {\cal R} \cdot A.
\ee
 The curvature matrix
\be
\label{k21}
 {\cal R}^i_j \stackrel{\rm def.}{=} d \varpi^i_j +\varpi^i_r  \wedge \varpi^r_j = \frac{1}{2} R^i_{jkl} dq^k \wedge dq^l,
\ee
where
\be
\label{k22}
R^i_{jkl} \stackrel{\rm def.}{=}  \frac{\partial \Gamma^i_{jl}}{\partial q^k} 
-\frac{\partial \Gamma^i_{jk}}{\partial q^l} + \Gamma^i_{rl} \Gamma^r_{jk} - \Gamma^i_{rk} \Gamma^r_{jl}
\ee 
is a curvature tensor of the connection determined by coefficients $\Gamma^i_{jk}.$

At the end of this introduction devoted to the general theory of connection in fibre bundles we remind procedures of defining induced connections on tensor product of bundles, direct sum of bundles and a dual bundle.

\begin{de}
Assume that  two exterior covariant derivatives  (denoted by the same symbol $\partial$) in two vector bundles $E_1$ and $E_2$ are given. The {\bf induced exterior covariant derivatives} on $E_1 \otimes E_2$ and $E_1 \oplus E_2$ are determined  by rules 
\be
\label{k10}
\partial(v_1 \otimes v_2)= \partial v_1 \otimes v_2 + v_1 \otimes \partial v_2,
\ee 
\be
\label{k11}
\partial(v_1 \oplus v_2)= \partial v_1 \oplus \partial v_2
\ee
for $v_1 \in E_1, v_2 \in E_2.$
\end{de}
\begin{de}
Suppose $v \in C^{\infty}(E), v^* \in C^{\infty}(E^*)$ and the pairing \\ $<v,v^*>\: \in C^{\infty}(M).$
The {\bf induced exterior covariant derivative}  on $E^*$  is determined by the relation
\be
\label{k20}
d <v,v^*>= <\partial v,v^*>+<v,\partial v^*>.
\ee
\end{de}
Relations (\ref{k10})--(\ref{k20}) contain of course the recipe  for  building {\bf induced connections} on $E_1 \otimes E_2, E_1 \oplus E_2 $ and $E^*$ respectively. 
 \vspace{0.5cm}

After that introduction we come back to the Weyl bundle and 
 present constructions of two connections: symplectic and abelian one on the Weyl bundle ${\cal P^*M}[[\hbar]].$ Both of those connections play crucial role in the definition of the $*$-product on a symplectic manifold ${\cal M}.$ As before we assume that $\dim {\cal M}=2n. $

From the Darboux theorem (see \cite{cannas}) for each point ${\tt p}$ on a symplectic manifold $  {\cal M}$ there exists a chart $({\cal U}_z, \phi_z)$ such that ${\tt p} \in {\cal U}_z$ and on $ {\cal U}_z$  in local coordinates  $(q^1, \ldots, q^{2n})$ determined by $\phi_z$  the symplectic form equals
\[
\omega= dq^{n+1} \wedge dq^1 + \ldots + dq^{2n} \wedge dq^n.
\]
A chart $({\cal U}_z, \phi_z)$ is called a {\bf Darboux  chart}. A set of $C^{\infty}$ compatible Darboux charts covering the whole manifold ${\cal A}= \{({\cal U}_z,\phi_z)\}_{z \in J}$ constitutes a {\bf Darboux atlas} on ${\cal M}.$

Suppose that on the tangent bundle ${\cal TM}$ a torsion-free connection is done. This connection $\varpi$ is locally determined by  sets of smooth real functions $\Gamma_{ij}^k\;, \; \\ 1 \leq i,j,k \leq 2n. $ Using formulas (\ref{k20}) and (\ref{k11}) we extend $\varpi$ easily on the bundle ${\cal T^*M} \otimes {\cal T^*M}.$
\begin{de} 
A torsion-free connection $\varpi $ on the symplectic manifold ${\cal M}$ is called {\bf symplectic}
if at each point of ${\cal M}$ 
\be
\label{doda0}
\partial \omega =0. 
\ee
\end{de}
It can  be proved \cite{my2} that each symplectic manifold may be endowed with some symplectic connection.  In a Darboux chart coefficients of symplectic connection
\be
\label{x1}
\Gamma_{ijk} \stackrel{\rm def.}{=}\omega_{li}\Gamma^{l}_{jk} 
\ee
are symmetric in all the indices $\{i,j,k\}.$ The matrix of symplectic connection 
\[
\varpi^i_j= \sum_{k=1}^{2n} \Gamma^i_{jk}dq^k.
\] 
A symplectic manifold equipped with a symplectic connection is often called a Fedosov manifold.

Let us consider the collection
\be
\label{k23}
({\cal T^*M})^l \stackrel{\rm def.}{= }\bigcup_{{\tt p} \in {\cal M}} (T^*_{\tt p}M)^l
\ee 
of spaces $(T^*_{\tt p}M)^l$ taken at all of points of the symplectic manifold ${\cal M}.$ It is easy to notice that this collection is a vector bundle which we will denote by $(({\cal T^*M})^l, \pi, {\cal M}).$ For $l \geq 1$
 we are able to introduce a local frame field on $(({\cal T^*M})^l, \pi, {\cal M})$
\newline
 $ \tilde{\bf e}_{\bf 1} \stackrel{\rm def.}{=} \underbrace{dq^1 \odot dq^1 \odot \ldots \odot dq^1}_{\rm l-times} ,$
\newline
$ \tilde{\bf e}_{\bf _2} \stackrel{\rm def.}{=} \underbrace{dq^1 \odot dq^1 \odot \ldots \odot dq^2}_{\rm l-times} ,$
\newline
$\vdots$
\newline
$
\tilde{\bf e}_{\bf \frac{(2n+l-1)!}{l!(2n-1)!}} \stackrel{\rm def.}{=} \underbrace{dq^{2n} \odot dq^{2n} \odot \ldots \odot dq^{2n}}_{\rm l-times}. $
\newline
The relation between indices $i_1, \ldots, i_{l}$ in the symmetric tensor product \\ $\tilde{\bf e}_{\bf k}=dq^{i_1} \odot dq^{i_2} \odot \ldots \odot dq^{i_l} $ and the number $k$ is given by the formula
\be
\label{k24}
k=\left( \begin{array}{c}
2n+l-1 \\ l
\end{array} \right) - \sum_{s=1}^{l}
\left( \begin{array}{c}
2n+s-i_{l-s+1}-1 \\ s
\end{array} \right). 
\ee
The proof of this fact is given in Appendix.

We look for the matrix of the induced connection on $({\cal T^*M})^l.$ 
From (\ref{k10})
\[
\partial(dq^{i_1} \odot dq^{i_2} \odot \ldots \odot dq^{i_l} )= \sum_{j=1}^{2n}
\sum_{r=1}^l -\varpi^j_{i_r} dq^{j } \odot dq^{i_1} \odot \ldots \odot \check{dq^{i_r}} \odot \ldots \odot dq^{i_l}.
\]
As usually the symbol $\check{dq^{i_r}} $ denotes the omitted element.

Let us consider the
  $k$-th row of a matrix $\;_{l}\varpi$ representing the induced connection on $({\cal T^*M})^l. $ We conclude that  different from $0$ may be  only terms $\;_{l}\varpi^m_k$  such that   among $l$ parametres $i_r$ determining (via \ref{k24}) number $m$  at least $(l-1)$  has been taken from the set $i_1, \ldots, i_l.$ 
Matrix  $\;_{l}\varpi$ is $\frac{(2n+l-1)!}{l!(2n-1)!} \times \frac{(2n+l-1)!}{l!(2n-1)!}$ dimensional.

\underline{The Example }
\newline
 Given a connection matrix $\varpi$ on the tangent bundle ${\cal TM}$ assuming that  $n=1, l=2$. The local frame system of the bundle $({\cal T^*M})^2$ contains three elements: \\ $dq^1 \odot dq^1, dq^1 \odot dq^2 $ and $dq^2 \odot dq^2.$ The matrix of induced connection $\;_{2}\varpi$ on $({\cal T^*M})^2$ looks like
\[
\:_2\varpi= \left[
\begin{array}{ccc}
-2 \varpi^1_1 & -2 \varpi^2_1 & 0 \\
-\varpi^1_2 & -\varpi^1_1 - \varpi^2_2 & -\varpi^2_1 \\
0 & -2 \varpi^2_1 & -2 \varpi^2_2
\end{array}
\right].
\]
\vspace{0.25cm}

In the third section analysing the construction of the Weyl vector space we showed that the order of elements in the formula (\ref{to5}) is determined by the degree, the power of $\hbar,$ indices of tensors and finally the real or imaginary nature of the element. Using those facts we represent that the Weyl algebra bundle as a double direct sum 
\be
\label{doda1}
{\cal P^*M}[[\hbar]]= \bigoplus_{z=0}^{\infty} \bigoplus_{k=0}^{[\frac{z}{2}]} \left( ({\cal T^*M})^{z-2k} \oplus ({\cal T^*M})^{z-2k} \right).
\ee
The local frame field of the Weyl bundle ${\cal P^*M}[[\hbar]]$ on ${\cal U} \subset {\cal M}$ contains
\newline
$ {\bf e_1} \stackrel{\rm def.}{=} \hat{i} \oplus \theta \oplus \theta \oplus \ldots,$
\newline
$
{\bf e_2} \stackrel{\rm def.}{=} \theta \oplus \hat{i} \oplus \theta \oplus \ldots, $
\newline
${\bf e_3} \stackrel{\rm def.}{=} \theta \oplus \theta \oplus  dq^1 \oplus \theta \oplus \ldots, $
\newline
${\bf e_4} \stackrel{\rm def.}{=} \theta \oplus \theta  \oplus  \theta  \oplus  dq^1 \oplus  \ldots, $
\newline
$\vdots$

Relation between an  index $a$ in ${\bf e_a}$ and indices $k,i_i, \ldots, i_l$ is expressed by  formulas (\ref{app8}) and
  (\ref{app9}) from Appendix. By $\hat{i}$ we understand a versor with the unitary length on a real line ${\mathbb R}.$  
The letter $\theta$ denotes the neutral
element in the vector space ${\mathbb R}.$

In the local frame field for a fixed even $z$ the matrix of induced connection
\[
\stackrel{{\bf z}} {\varpi} = \left[
\begin{array}{cccccc}
 \;_z \varpi & {\bf 0} & \ldots & {\bf 0}& {\bf 0} & {\bf 0} \\
 {\bf 0} & \;_z \varpi & \ldots & {\bf 0}& {\bf 0} & {\bf 0} \\
\vdots& \vdots& \ddots & \vdots& \vdots & \vdots \\
{\bf 0}& {\bf 0}& \ldots & \;_1 \varpi & {\bf 0} & {\bf 0} \\
{\bf 0}& {\bf 0}& \ldots & {\bf 0} & {\bf 0} & {\bf 0} \\
{\bf 0}& {\bf 0}& \ldots & {\bf 0} & {\bf 0} & {\bf 0} 
\end{array}
\right]
\]
or, for an odd $z$
\[
\stackrel{{\bf z}}{ \varpi } = \left[
\begin{array}{cccc}
\;_z \varpi & {\bf 0} & \ldots & {\bf 0} \\
 {\bf 0} & \;_z \varpi & \ldots & {\bf 0} \\
\vdots& \vdots& \ddots & \vdots \\
{\bf 0}& {\bf 0}& \ldots & \;_1 \varpi 
\end{array}
\right].
\]
Each matrix $\stackrel{{\bf z}}{ \varpi }$ is a square even dimensional matrix.

The matrix of induced connection on the Weyl bundle
\be
\label{ania}
 \varpi = \left[
\begin{array}{cccc}
{\bf 0}& {\bf 0}&  {\bf 0} &  \ldots \\
{\bf 0}& \stackrel{{\bf 1}}{ \varpi } &  {\bf 0} &  \ldots \\
{\bf 0}& {\bf 0} &  \stackrel{{\bf 2}}{ \varpi } &  \ldots \\
\vdots& \vdots & \vdots & \ddots  
\end{array}
\right].
\ee

In Darboux coordinates coefficients of the symplectic conection (\ref{x1}) are symmetric in all indices, so we may express the exterior covariant derivative in terms of the $\circ$-product. Indeed,
let us introduce a  connection $1$-form  as an element of \\ $C^{\infty}(({\cal T^*M})^2 \otimes \Lambda^1({\cal M}))$
\[
\Gamma= \Gamma[0,2]_{ij,k}dq^k \stackrel{\rm def.}{=}\Gamma_{ij,k}dq^k. 
\]
Moreover,
\begin{de} 
\label{def1}
The { commutator} of two smooth sections $u \in C^{\infty}( {\cal P^*M}[[\hbar]] \otimes \Lambda^{j_1}({\cal M}))$ and  $v \in 
C^{\infty}({\cal P^*M}[[\hbar]] \otimes \Lambda^{j_2}({\cal M}))$ is a smooth section of 
$ C^{\infty}({\cal P^*M}[[\hbar]] \otimes \Lambda^{j_1 + j_2}({\cal M}))$ such that
\be
\label{kom}
[u,v] \stackrel{\rm def.}{=}  u \circ v -(-1)^{j_1 \cdot j_2}v \circ u.
\ee
\end{de}
In further considerations we
 simplify a bit the notation we put ${\cal P^*M}[[\hbar]]  \Lambda({\cal M})$ instead of $\sum_{j=1}^{2n} {\cal P^*M}[[\hbar]] \otimes \Lambda^j({\cal M}).$

\begin{tw}
In Darboux coordinates
the exterior covariant derivative  $\partial v$ of  \\ $v \in C^{\infty} ({\cal P^*M}[[\hbar]]  \Lambda({\cal M}))$ reads
\be
\label{kon1}
\partial v = dv + \frac{1}{i \hbar}[\Gamma,v].
\ee
\end{tw}
The reason why we restrict ourselfs to Darboux charts is obvious - only in such charts connection is represented by a form with values in the Weyl algebra.
  
 In Fedosov paper \cite{6}
the above relation is just a definition of the exterior covariant derivative  in the Weyl bundle. We showed its geometrical
origin.

From the fact that the exterior covariant derivative  in the Weyl bundle can be put into frames of the algebra structure, we deduce that also curvature of the symplectic connection may be defined in terms of ${\cal P^*M}[[\hbar]] \Lambda({\cal M}).$ Indeed, 
curvature of the symplectic connection $\Gamma$ is a smooth section  $C^{\infty}({\cal P^*M}[[\hbar]]\otimes \Lambda^2({\cal M}))$ and can be written as 
\be
\label{doda2}
{\cal R} = d \Gamma + \frac{1}{i \hbar} \Gamma \circ \Gamma.
\ee
The action of ${\cal R}$ on a smooth section $X$ of a tangent bundle ${\cal TM}$ is given by
\[
{\cal R}(X)= \frac{1}{4} R_{ijkl}X^i X^j dq^k \wedge dq^l,
\]
where 
\be
\label{doda4}
R_{ijkl}= \frac{\partial \Gamma_{ijl}}{\partial q^k} -
 \frac{\partial \Gamma_{ijk}}{\partial q^l} + \omega^{zu}\Gamma_{zil}\Gamma_{ujk}-
\omega^{zu}\Gamma_{zik}\Gamma_{ujl}.
\ee
Thus the second exterior covariant derivative of $v \in C^{\infty} ({\cal P^*M}[[\hbar]]  \Lambda({\cal M}))$ is described by a formula
\be
\label{doda5}
\partial^2 v = \frac{1}{i \hbar}[{\cal R},v].
\ee

%%%%%%%%%%%%%%%%%%%%%%%%%%%%%%%%%%%%%%%%%%%%%%%%%%%%%%%%%%%%%%%%%%%%%%%%%%%%%%%%%%%%%%%%%%%%%%%%%%%%%%%%%%%%%%%%%%%%%%%%%%%
%%%%%%%%%%%%%%%%%%%%%%%%%%%%%%%%%%%%%%%%%%%%%%%%%%%%%%%%%%%%%%%%%%%%%%%%%%%%%%%%%%%%%%%%%%%%%%%%%%%%%%%%%%%%%%%%%%%%%%%%%%%

\section{The $*$-product on a symplectic manifold ${\cal M}$}

Crucial role in deformation  quantization is played by a noncommutative nonabelian $*$-product being a counterpart of the product of operators from Hilbert space formulation of quantum theory. There is no unique way to introduce this product in case when the phase space of the system is different from ${\mathbb R}^{2n}.$ 
In this section we present briefly the Fedosov construction of the $*$-product on a symplectic manifold $({\cal M}, \omega).$  We concentrate on a geometric aspect of the problem omitting all of technical proofes.

There exists infinitely many different connections in the bundle ${\cal P^*M}[[\hbar]]  \Lambda({\cal M}).$ Especially important to our purposes is one of so called  abelian connections $\tilde{\Gamma}$. 
\begin{de}
A connection $\tilde{\Gamma}$ is called {\bf abelian}, if its  curvature $\Omega$ is a $2$-form with values in the bundle $\bigoplus_{k=0}^{\infty}\left(({\cal T^*M})^0_k \oplus ({\cal T^*M})^0_k \right) \otimes \Lambda^2({\cal M}).$ 
\end{de}
In the definition we put $ ({\cal T^*M})^0_k = ({\cal T^*M})^0$ for every $k.$

Since that  for each $v \in C^{\infty} ({\cal P^*M}[[\hbar]]  \Lambda({\cal M}))$ the second exterior covariant derivative $D$ determined by $\tilde{\Gamma}$ equals (compare (\ref{doda5}))
\be
D^2 v = \frac{1}{i\hbar}[\Omega,v]=0.
\ee
From among the set of abelian connections especially important for physical applications is the one of the form
\be
\label{nieoc}
\tilde{\Gamma}(X)\stackrel{\rm def.}{=}   \omega_{i,j}X^i dq^j + \Gamma_{i_1 i_2,j}X^{i_1}X^{i_2}dq^j  + r.
\ee
The term $r \in {\cal P^*M}[[\hbar]] \otimes \Lambda^{1}({\cal M}), \; \deg(r) \geq 3$ is determined by some recurrential formula (for details see \cite{6}, \cite{7}) and it depends only on the symplectic curvature ${\cal R}.$ 
In this case 
\[
\Omega= - \frac{1}{2}\omega_{j_1 j_2}dq^{j_1} \wedge dq^{j_2}.
\]
 To avoid confusion which indices are of the Weyl bundle and which ones of a differential form  we put in (\ref{nieoc}) the result of acting by $\tilde{\Gamma}$ on some arbitrary fixed vector field from the tangent bundle ${\cal TM}.$

It is easy to notice that the component of the abelian connection (\ref{nieoc}) with the degree $z$ acting on the form $v \in {\cal P^*M}[[\hbar]] \Lambda({\cal M})$ gives as a result the element \\ $u \in {\cal P^*M}[[\hbar]] \Lambda({\cal M}) $ for which 
\[
\deg(u)= \deg(v) + z-2.
\]
Since that denoting by $\tilde{\varpi}^{(\deg(v))}_{(\deg(u))}$ a part of the connection matrix $\varpi$ appearing as the result of presence of the abelian connection $\tilde{\Gamma}$ we see that 
\be
\tilde{\varpi} = \left[
\begin{array}{cccc}
{\bf 0}& {\bf 0}&  {\bf 0} &  \ldots \\
\tilde{\varpi}^{(2)}_{(1)}& \stackrel{{\bf 1}}{ \varpi } &  \tilde{\varpi}^{(2)}_{(3)} &  \ldots \\
{\bf 0}& \tilde{\varpi}^{(3)}_{(2)} &  \stackrel{{\bf 2}}{ \varpi } &  \ldots \\
\vdots& \vdots & \vdots & \ddots  
\end{array}
\right]
\ee
(compare (\ref{ania})). Notice that the abelian connection $\tilde{\Gamma}$ mixes up tensors of different ranges and terms standing at different powers of the deformation parameter $\hbar.$

What is interesting, the set of $0$-forms belonging to  ${\cal P^*M}[[\hbar]] \otimes \Lambda^{0}({\cal M})$ such that $Dv=0,$ constitutes the subalgebra of the Weyl algebra ${\cal P^*M}[[\hbar]] .$ 
We will denote this subalgebra by ${\cal P^*M}_D[[\hbar]] .$ 
\begin{de}
A {\bf projection} 
\[
\sigma:{\cal P^*M}_D[[\hbar]] \otimes \Lambda^{0}({\cal M}) \longrightarrow \bigoplus_{k=0}^{\infty} \left(
({\cal T^*M})^0_k \oplus ({\cal T^*M})^0_k  \right) 
\]
 assigns to each  section $v \in {\cal P^*M}_D[[\hbar]] \otimes \Lambda^0({\cal M}) $ its part $\sigma(v) \in \bigoplus_{k=0}^{\infty} \left(
({\cal T^*M})^0_k \oplus ({\cal T^*M})^0_k  \right). $ 
\end{de}
It has been proved (see \cite{6}, \cite{7}) that for each $ f \in C^{\infty}  \left( \bigoplus_{k=0}^{\infty} \left(
({\cal T^*M})^0_k \oplus ({\cal T^*M})^0_k  \right) \right)$ there exists a {\bf unique} smooth section $v \in C^{\infty}({\cal P^*M}_D[[\hbar]] \otimes \Lambda^0({\cal M}))$ such that \\ $\sigma(v)=f.$ 

\vspace{0.5cm}
We arrived to the point crucial for physical applications of the mathematical machinery presented above.
\begin{de} 
Let $f_1, f_2$ be two  smooth sections of $C^{\infty}  \left( \bigoplus_{k=0}^{\infty} \left(
({\cal T^*M})^0_k \oplus ({\cal T^*M})^0_k  \right) \right)$ . 
The $*$-{product} of them is defined as 
\be \label{koniec}
f_1 * f_2 \stackrel{\rm def.}{=}
\sigma\left( \sigma^{-1}(f_1) \circ \sigma^{-1}(f_2) \right). \ee
\end{de}
This $*$-product defined here can be considered  as a generalization of the
Moyal product of  Weyl type defined for   $ M = {\mathbb R}^{2n}$. 
It has the following properties:
\renewcommand{\theenumi}{\bf \arabic{enumi}}
\begin{enumerate}
\item The definition of $*$-product is invariant under Darboux transformations. Since that we can multiply functions on an arbitrary symplectic manifold. In case of the Moyal product of the Weyl type we are restricted to the symplectic manifold ${\mathbb R}^{2n}$ and Darboux coordinates with vanishing $1$-form of a symplectic connection $\Gamma=0.$ In fact we could extend the definition of $*$-product on an arbitrary atlas on the symplectic manifold ${\cal M},$ but in such case it would be necessary to modify recurrential formulas for  the term $r$ in (\ref{nieoc}) and for $\sigma^{-1}.$ 
 
\item In the limit $\hbar \rightarrow 0^+ $ the $*$-product of $f_1, f_2 \in C^{\infty}\left(({\cal T^*M})^0_0 \oplus ({\cal T^*M})^0_0 \right)$  turns
into the commutative point-wise multiplication of functions, i.e.
\be\label{limite}
 \lim_{\hbar \rightarrow 0^+ }f_1 * f_2= f_1 \cdot f_2. 
\ee
This relation expresses the fact that the classical mechanics is a limit of quantum physics for the Planck constant tending to $0^+.$
 \item
 The multiplication (\ref{koniec}) is associative but noncommutative. The noncommutativity of two smooth sections $f_1, f_2 \in C^{\infty}  \left( \bigoplus_{k=0}^{\infty} \left(
({\cal T^*M})^0_k \oplus ({\cal T^*M})^0_k  \right) \right)$ is measured by their Moyal bracket
\[
\{f_1,f_2\}_M \stackrel{\rm def.}{= }\frac{1}{i \hbar}(f_1*f_2-f_2*f_1)
\]
which plays a role of commutator from standard formulation of quantum mechanics on Hilbert space.

\item 
When $  {\cal M} = {\mathbb R}^{2n}$ the product defined
above is just the Moyal product of
 Weyl type. This fact confirms consistency of the Fedosov approach with the best known case of quantum deformation. 
\end{enumerate}
Some computations done with the $*$-product (\ref{koniec}) can be found in \cite{my2}.

\vspace{0.5cm}

{\bf Acknowledgments}

I am grateful to Prof. Manuel Gadella from the Valladolid University for indicating difficulties in construction of infinite dimensional spaces. I also thank Ms Magda Nockowska from the Mathematical Department of Technical University of Lodz for fruitful discussion.  

\renewcommand{\theequation} {\Alph{section}.\arabic{equation}}
\setcounter{section}{1}
\section*{Appendix  \\Position of the element $v[k,l]_{i_1 \ldots i_l}$ in the series (\ref{to5})}

In this Appendix we shall explain the relation between indices $k,i_1, \ldots i_l$ of an element  $v[k,l]_{i_1 \ldots i_l}$ of the Weyl algebra  and its position in the series
\[
\phi(v)=(Re(v[0,0]),Im(v[0,0]),Re(v[0,1]_{1}),Im(v[0,1]_{1}),\ldots ) 
\]
representing a vector $v$ of the Weyl space $P^*_{\tt p}M[[ \hbar]].$ 
Formulas proved in this Appendix are also useful in further considerations on the construction of a local frame field in the Weyl bundle.
 
We will achieve our aim in four steps.
\renewcommand{\theenumi}{\arabic{enumi}}
\begin{enumerate}
\item
At the beginning we find the place $P_1$ of $v[k,l]_{i_1 \ldots i_l}$ among elements with the same $k$ and $l.$
\item
After that we consider the relation between a power $k$ of $\hbar$ and the position $P_2$  for elements with the same degree $2k+l.$
\item
 In the third step we include the influence of the degree on the position $P_3.$
\item
Finally we present the complete formula expressing the relation between the real and imaginary part of $v[k,l]_{i_1 \ldots i_l}$ and its position ${\bf P}$ in (\ref{to5}).
\end{enumerate}

Let us start from the case when $k$ is fixed and $l \geq 1.$ We look for the position  of 
$v[k,l]_{i_1 \ldots i_l}$ in the series of real elements parametrizing by the same $k$ and $l.$ Elements are ordered according to the rule valid for the relation (\ref{za1}). 
The position $P_1(v[k,l]_{i_1\ldots i_l})$ of $v[k,l]_{i_1 \ldots i_l}$ in that series equals
\[
\sum_{z_1=1}^{i_1-1} \sum_{z_2=z_1}^{2n} \ldots \sum_{z_l=z_{l-1}}^{2n} 1 +
\sum_{z_2=i_1}^{i_2-1} \sum_{z_3=z_2}^{2n} \ldots \sum_{z_l=z_{l-1}}^{2n} 1 + \cdots
+ \sum_{z_{l-1}=i_{l-2}}^{i_{l-1}-1}\sum_{z_l=z_{l-1}}^{2n}1 + \sum_{z_l=i_{l-1}}^{i_l}1=
\]
\be
\label{app2}
\sum_{z_1=1}^{i_1-1} \sum_{z_2=z_1}^{2n} \ldots \sum_{z_l=z_{l-1}}^{2n} 1 +
\sum_{z_2=1}^{i_2-i_1} \sum_{z_3=z_2}^{2n-i_1+1} \ldots \sum_{z_l=z_{l-1}}^{2n-i_1+1} 1 + \cdots
+ \sum_{z_{l-1}=1}^{i_{l-1}-i_{l-2}}\sum_{z_l=z_{l-1}}^{2n-i_{l-2}+1}1 + \sum_{z_l=1}^{i_l-i_{l-1}+1}1.
\ee
Let us consider the following expression
\[
\sum_{z_1=1}^{m} \sum_{z_2=z_1}^{m} \ldots \sum_{z_l=z_{l-1}}^{m} 1 =
\sum_{z_1=1}^{a} \sum_{z_2=z_1}^{m} \ldots \sum_{z_l=z_{l-1}}^{m} 1 +
\sum_{z_1=a+1}^{m} \sum_{z_2=z_1}^{m} \ldots \sum_{z_l=z_{l-1}}^{m}1= 
\]
\[
\sum_{z_1=1}^{a} \sum_{z_2=z_1}^{m} \ldots \sum_{z_l=z_{l-1}}^{m} 1 +
\sum_{z_1=1}^{m-a} \sum_{z_2=z_1}^{m-a} \ldots \sum_{z_l=z_{l-1}}^{m-a}1 \stackrel{\rm (\ref{dziwo})}{=} 
\]
\[
\sum_{z_1=1}^{a} \sum_{z_2=z_1}^{m} \ldots \sum_{z_l=z_{l-1}}^{m} 1 +
\frac{(m+l-a-1)!}{l!(m-a-1)!}.
 \]
But from (\ref{dziwo}) we know that $\sum_{z_1=1}^{m} \sum_{z_2=z_1}^{m} \ldots \sum_{z_l=z_{l-1}}^{m} 1 = \frac{(m+l-1)!}{l!(m-1)!} 
$ so
\be
\label{app1}
\sum_{z_1=1}^{a} \sum_{z_2=z_1}^{m} \ldots \sum_{z_l=z_{l-1}}^{m} 1= \frac{(m+l-1)!}{l!(m-1)!} - \frac{(m+l-a-1)!}{l!(m-a-1)!}.
\ee
Relation (\ref{app1}) works also for $l=1.$ In this case the right-hand side of (\ref{app1}) is independent of $m$ and we see that $\sum_{z_1=1}^{a}=a.$
Substituting (\ref{app1}) into (\ref{app2}) we have
\[
P_1(v[k,l]_{i_1 \ldots i_l})=\frac{(2n+l-1)!}{l!(2n-1)!}-\frac{(2n+l-i_1)!}{l!(2n-i_1)!} + \frac{(2n+l-i_1-1)!}{(l-1)!(2n-i_1)!}-
\frac{(2n+l-i_2-1)!}{(l-1)!(2n-i_2)!}+ \ldots
\]
\[
+  \frac{(2n-i_{l-2}+2)!}{2!(2n-i_{l-2})!}         -  \frac{(2n-i_{l-1}+2)!}{2!(2n-i_{l-1})!}        +(i_l-i_{l-1}+1).
\]
Putting together terms containing $i_{l_r}$ we see that 
\[
P_1(v[k,l]_{i_1 \ldots i_l})=
\frac{(2n+l-1)!}{l!(2n-1)!}- \frac{(2n+l-i_1-1)!}{l!(2n-i_1-1)!}- \ldots - \frac{(2n-i_{l-1}+1)!}{2!(2n-i_{l-1}-1)!}-
\frac{(2n-i_l)!}{1!(2n-i_l-1)!}.
\]
Notice that although 
\[
- \frac{(2n-i_{l-1}+1)!}{2!(2n-i_{l-1}-1)!}= -  \frac{(2n-i_{l-1}+2)!}{2!(2n-i_{l-1})!}        -i_{l-1}+1 +{\bf 2n}
\]
and
\[
- \frac{(2n-i_l)!}{1!(2n-i_l-1)!}= i_l - {\bf 2n}
\]
the sum
\[
- \frac{(2n-i_{l-1}+1)!}{2!(2n-i_{l-1}-1)!} - \frac{(2n-i_l)!}{1!(2n-i_l-1)!}= -  \frac{(2n-i_{l-1}+2)!}{2!(2n-i_{l-1})!}        -i_{l-1}+1 + i_l
\]
as  is required.

Finally we may write  
\be
\label{app3}
P_1(v[k,l]_{i_1 \ldots i_l})=\left( \begin{array}{c}
2n+l-1 \\ l
\end{array} \right) - \sum_{s=1}^{l}
\left( \begin{array}{c}
2n+s-i_{l-s+1}-1 \\ s
\end{array} \right). 
\ee
Notice that the above  relation is true also for scalars i.e. when $l=0.$
 
In the next step we take into account the degree of $v[k,l]_{i_1\ldots i_l}.$
 From the definition \ref{dedegree} we see that ${\rm deg}(v[k,l]_{i_1 \ldots i_l})=2k+l. $ We  denote it by $d.$ The position $ P_2(v[k,l]_{i_1\ldots i_l})$ of $v[k,l]_{i_1\ldots i_l}$ among elements with the same degree is given by
\[
 P_2(v[k,l]_{i_1\ldots i_l})=\sum_{g=0}^{k-1}
\left( \begin{array}{c}
2n+d-2g-1 \\ d-2g
\end{array} \right) + P_1(v[k,l]_{i_1\ldots i_l}) = 
\]
\[
= -\left( \begin{array}{c}
2n+l-1 \\ l
\end{array} \right) \:_3 F_2(\{1, \frac{1}{2} - \frac{l}{2},- \frac{l}{2} \}, \{\frac{1}{2} - \frac{l}{2}-n, 1-\frac{l}{2}-n  \},1)+
\]
\be
\label{app4}
  \left( \begin{array}{c}
2n+2k+l-1 \\ 2k+l
\end{array} \right) 
\:_3 F_2(\{1, -k - \frac{l}{2},\frac{1}{2} - \frac{l}{2} -k\}, \{\frac{1}{2} - \frac{l}{2}-k -n, 1-\frac{l}{2}-k-n  \},1) 
+ P_1(v[k,l]_{i_1\ldots i_l}) 
\ee 
By $\:_3 F_2(\{a_1,a_2,a_3\},\{b_1,b_2\},x)$ we denote the generalized hypergeometric function (see \cite{ryz}). 
 From (\ref{app4}) we obtain that
the total number of real elements of the same degree $d$ is
\be
\label{app7}
\sum_{g=0}^{[\frac{d}{2}]}
\left( \begin{array}{c}
2n+d-2g-1 \\ d-2g
\end{array} \right)= 
\left( \begin{array}{c}
2n+d-1 \\ d
\end{array} \right)\:_3F_2(\{1,\frac{1}{2}-\frac{d}{2},-\frac{d}{2}\},\{\frac{1}{2}- \frac{d}{2}-n, 1 -\frac{d}{2}-n\},1).  
\ee
Since  we conclude that  
\[
 P_3(v[k,l]_{i_1 \ldots i_l})= P_2(v[k,l]_{i_1\ldots i_l})+  
\]
\[
 \sum_{c=0}^{d-1}
\left( \begin{array}{c}
2n+c-1 \\ c
\end{array} \right)\:_3F_2(\{1,\frac{1}{2}-\frac{c}{2},-\frac{c}{2}\},\{\frac{1}{2}- \frac{c}{2}-n, 1 -\frac{c}{2}-n\},1)=  
\]
\be
P_2(v[k,l]_{i_1\ldots i_l})+
  \left\{
\begin{array}{r}
-\frac{1}{4^{n+1}}- \left( \begin{array}{c}
2n+d+1 \\ d+1
\end{array} \right)
\;_3F_2(\{1,1+\frac{d}{2}+n,\frac{3}{2}+\frac{d}{2}+n\},\{1+\frac{d}{2},\frac{3}{2}+\frac{d}{2}\},1)  \\
\;  {\rm for \;\; even \;\; d} \\
\vspace{0.25cm}\\
-\frac{1}{4^{n+1}}- \left( \begin{array}{c}
2n+d \\ d
\end{array} \right)
\;_3F_2(\{1,\frac{1}{2}+\frac{d}{2}+n,1+\frac{d}{2}+n\},\{\frac{1}{2}+\frac{d}{2},1+\frac{d}{2}\},1)  \\
+ \left( \begin{array}{c}
2n+d-2 \\ d-1
\end{array} \right)
\;_3F_2(\{1,\frac{1}{2}-\frac{d}{2},1-\frac{d}{2}\},\{1-\frac{d}{2}-n,\frac{3}{2}-\frac{d}{2}-n\},1)  \\
\;  {\rm for \;\; odd \;\; d.}
 \end{array} 
\right.
\ee
Finally we arrive at the result that the position of the element $v[k,l]_{i_1\ldots i_l}$ in the series (\ref{to5}) is described by two relations
\be
\label{app8}
{\bf P}(Re(v[k,l]_{i_1 \ldots i_l}))= 2  P_3(v[k,l]_{i_1\ldots i_l})-1 
\ee 
and
\be
\label{app9}
{\bf P}(Im(v[k,l]_{i_1\ldots i_l}))= 2  P_3(v[k,l]_{i_1\ldots i_l}). 
\ee 
Mappings (\ref{app8}) and (\ref{app9}) are one-to-one so from the position of the element in the series (\ref{to5}) it is possible to reconstruct the  sequence $k,i_1, \ldots, i_l.$

\end{document}